# Realistic Derivation of Heisenberg Dynamics

Paul J. Werbos[*]
National Science Foundation[†], Room 675
Arlington, VA 22203

ABSTRACT

Einstein conjectured long ago that much of quantum mechanics might be derived as a statistical formalism describing the dynamics of classical systems. Bell's Theorem experiments have ruled out complete equivalence between quantum field theory (QFT) and classical field theory (CFT), but an equivalence between dynamics is not only possible but provable in simple bosonic systems. Future extensions of these results might possibly be useful in developing provably finite variations of the standard model of physics.

PAC codes: 3.65.Ta, 11.10, 05.20.Gg, 03.65.Ca

## I. INTRODUCTION AND SUMMARY

The goal of this letter is to present a new formalism for analyzing the statistical dynamics of "classical fields" – of ODE or PDE derived from a classical Lagrangian density of the usual Hamiltonian form. To keep the notation simple, I will focus on the ODE case with:

$$L = \tfrac{1}{2} \sum_{j=1}^{n} (\dot{\varphi}_j \pi_j - \dot{\pi}_j \varphi_j) - H(\underline{\varphi}, \underline{\pi}) \tag{1}$$

where the state of the system at time t is defined by the two classical vectors $\boldsymbol{\varphi}, \boldsymbol{\pi} \in R^n$. The generalization to PDE is discussed in Appendix B.

The following section will show how a probability distribution or statistical ensemble of possible states, $Pr(\boldsymbol{\varphi}, \boldsymbol{\pi})$ may be "encoded" or "mapped" into a "classical density matrix" $\rho$, which is dual to the usual field operators of the Heisenberg formulation of QFT[1]. I will then define simple field operators $\Phi_j$ and $\Pi_j$ which have the properties that $Tr(\Phi_j\rho)=<\varphi_j>$, $Tr(\Pi_j\rho)=<\pi_j>$ and both obey the usual Heisenberg/Liouville dynamics as described by Weinberg[1]. Note that I use angle brackets to denote the expectation value of classical stochastic variables.

The concluding section will discuss the possible significance and future extension of these results.

## II. CONCEPTS AND DEFINITIONS

Efforts to understand the statistics of classical fields have a long history which cannot be reviewed comprehensively in a brief letter; however, some historical concepts are essential to understanding of the new work.

For centuries, statisticians have studied how probability distributions $Pr(\boldsymbol{\varphi})$ can be represented and studied in terms of statistical moments, correlations or cumulants. For example, the set of real numbers

$$u_{i_1,i_2,\ldots,i_n} = \left\langle \varphi_1^{i_1} \varphi_2^{i_2} \ldots \varphi_n^{i_n} \right\rangle \tag{2}$$

includes the statistical mean and (implicitly) the covariance, skewness and kurtosis, etc. The set of all such u values for a given probability distribution may be seen, mathematically, as a kind of a real wave function, as a real vector **u** in the usual Fock-Hilbert space of QFT!

Many classical physicists have tried to "close turbulence" by deriving linear or nonlinear dynamical equations for the statistical moments. However, when the vector **u** is used directly – without scaling and without exploiting its special properties – the dynamical equations tend to display a problem of infinite regress, explicitly or implicitly. Many of those efforts are analogous to older textbooks which try to solve simultaneous linear equations by writing each equation separately and manipulating them one by one; by shifting over to Fock space concepts from QFT, we can achieve a simplicity and power analogous to the power which matrix algebra provides in working with systems of linear equations.

In a previous paper, we[2] suggested that the statistical moments are slightly easier to analyze if we first scale them:

$$v_{i_1,i_2,...,i_n} = \frac{u_{i_1,i_2,...,i_n}}{\sqrt{i_1! i_2! ... i_n!}}, \tag{3}$$

which is equivalent to the more compact and modern definition:

$$\underline{v}(\underline{\varphi}) = \exp\left(\sum_{j=1}^{n} \varphi_j a_j^+\right) | 0 >, \tag{4}$$

where $a_j^+$ is the usual creation operator of QFT. This lets us represent a probability distribution $\Pr(\underline{\varphi})$ by:

$$\underline{v} = <\underline{v}(\underline{\varphi})> = \int \underline{v}(\underline{\varphi}) \Pr(\underline{\varphi}) d^n \underline{\varphi} \tag{5}$$

Equation 4 is very similar to the mappings used by Segal and by Rajeev[3] in the analysis of Bose-Fermi equivalences; however, it is not a special case, because the operator in the exponential is not antiHermitian,

It is easy to calculate $|\underline{v}|^2$ and verify that it is bounded but not normalized to 1 for most states of physical interest. Furthermore, $\underline{v}$ only represents a distribution $\Pr(\underline{\varphi})$, not a joint distribution $\Pr(\underline{\varphi},\underline{\pi})$. Therefore, to represent the statistical moments of the system given in equation 1, I propose the following normalized (complex) vector:

$$\underline{w}(\underline{\varphi},\underline{\pi}) = \exp\left(\sum_{j=1}^{n}\left((\varphi_j + i\pi_j)a_j^+ - \tfrac{1}{2}(\varphi_j^2 + \pi_j^2)\right)\right)|0\rangle \tag{6}$$

The classical density matrix is then defined by

$$\rho = \iint \underline{w}(\underline{\varphi},\underline{\pi})\underline{w}^H(\underline{\varphi},\underline{\pi})\Pr(\underline{\varphi},\underline{\pi})d^n\underline{\varphi}\,d^n\underline{\pi} \quad, \tag{7}$$

where the superscript H denotes the usual Hermitian conjugate. Vectors $\underline{w}$ which represent a specific state $(\underline{\varphi},\underline{\pi})$ as in equation 6 will be called "pure states." Matrices $\rho$ which represent an ensemble of states as in equation 7 will be called "physically realizable."

For this ODE system, the two basic field operators take on a simple form:

$$\Phi_j = \tfrac{1}{2}(a_j + a_j^+) \tag{8}$$

$$\Pi_j = \frac{1}{2i}(a_j - a_j^+) \tag{9}$$

III. PROPERTIES AND DYNAMICS

It is easy to verify that $\Phi_j$ and $\Pi_j$ possess the usual properties for field operators, apart from minor scaling factors, such as:

$$[\Phi_j, \Pi_k] = \delta_{jk}(\tfrac{1}{2})(\tfrac{1}{2i})(-2) = \tfrac{i}{2}\delta_{jk} \tag{10}$$

Likewise:

$$[\Phi_j, (\Pi_k)^m] = \tfrac{i}{2}\delta_{jk} m (\Pi_k)^{m-1} \tag{11}$$

For any analytic function of $\underline{\varphi}$ and $\underline{\pi}$,

$$f(\underline{\varphi},\underline{\pi}) = \sum_{i_1,\ldots,i_n, j_1,\ldots,j_n=0}^{\infty} C_{j_1,\ldots,j_n}^{i_1,\ldots,i_n} \varphi_1^{i_1}\varphi_2^{i_2}\ldots\varphi_n^{i_n}\pi_1^{j_1}\ldots\pi_n^{j_n}, \tag{12}$$

we may define two different "quantized" versions, a canonical version

$$f_c(\underline{\Phi},\underline{\Pi}) = \sum_{i_1,\ldots,i_n, j_1,\ldots,j_n=0}^{\infty} C_{j_1,\ldots,j_n}^{i_1,\ldots,i_n} \Phi_1^{i_1}\Phi_2^{i_2}\ldots\Phi_n^{i_n}\Pi_1^{j_1}\ldots\Pi_n^{j_n} \tag{13}$$

and a normal version

$$f_n(\underline{\Phi},\underline{\Pi}) = \; :f_c(\underline{\Phi},\underline{\Pi}): \tag{14}$$

where the colons are the usual notation for normal product. It is straightforward but tedious to prove the following extensions of equation 11:

$$[\Phi_j, f_n(\underline{\Phi},\underline{\Pi})] = \tfrac{i}{2}\left(\frac{\partial f}{\partial \pi_j}\right)_n \tag{15}$$

$$[\Pi_j, f_n(\underline{\Phi},\underline{\Pi})] = -\tfrac{i}{2}\left(\frac{\partial f}{\partial \varphi_j}\right)_n \tag{16}$$

and likewise for the (easier) case of $f_c$.

Somewhat more novel is the fact that:

$$\begin{aligned}a_i\,\underline{w}(\underline{\varphi},\underline{\pi}) &= [a_i, \exp(\sum(\varphi_j + i\pi_j)a_j^+ - \tfrac{1}{2}(\ldots))]|0\rangle + \exp(\ldots)a_i|0\rangle \\ &= (\varphi_i + i\pi_i)\underline{w}(\underline{\varphi},\underline{\pi}) + 0\end{aligned} \tag{17}$$

Thus:

$$Tr(\rho \Phi_i) = \tfrac{1}{2}Tr(\rho a_i + \rho a_i^+) = \tfrac{1}{2}<\underline{w}^H a_i \underline{w}> + \tfrac{1}{2}<(a_i \underline{w})^H \underline{w}>$$
$$= \tfrac{1}{2}<\varphi_i + i\pi_i> + \tfrac{1}{2}<\varphi_i - i\pi_i> = <\varphi_i>. \qquad (18)$$

and similarly for $\Pi_i$ and $\pi_i$. More generally, we may deduce that:

$$Tr(\rho f_n(\underline{\Phi},\underline{\Pi})) = <f(\underline{\varphi},\underline{\pi})> \qquad (19)$$

for any analytic function f. Intuitively, the effect of the normal product is to keep all the $a_i^+$ terms to the left, where they yield factors of ($\varphi_i$-i$\pi_i$), and to keep the $a_i$ terms to the right, where they yield factors of ($\varphi_i$+i$\pi_i$).

Finally, consider the classical dynamical equations for the system defined by equation 1:

$$\dot{\varphi}_j = \frac{\partial H}{\partial \pi_j} \qquad (20)$$

$$\dot{\pi}_j = -\frac{\partial H}{\partial \varphi_j} \qquad (21)$$

Taking the expectation value of equation 20, and exploiting equations 15 and 19, we may deduce:

$$<\dot{\varphi}_j> = -2i\,Tr(\rho[\Phi_j, H_n(\underline{\Phi},\underline{\Pi})]) \qquad (22)$$

Likewise:

$$<\dot{\pi}_j> = -2i\,Tr(\rho[\Pi_j, H_n(\underline{\Phi},\underline{\Pi})]) \qquad (23)$$

It is tempting to simply interpret equation 22 to mean:

$$"\dot{\Phi}_j" = -2i[\Phi_j, H] \qquad (24)$$

and declare that we are all done (aside from scalar factors which, like h=c=1 units, do not change the physics.) Note the implicit assumption – as in real-world QFT[1,4] – that the Hamiltonian operator H actually used by QFT is the normal-product Hamiltonian.

To be more rigorous, one must follow the procedures given by Weinberg[1]. The operators $\Phi_j$ and $\Pi_j$ above correspond to the $q_n(x,0)$ and $p_n(x,0)$ operators in his development. The true Heisenberg operators, $Q_n$ and $P_n$, are simply defined by his equations 7.1.28 and 7.1.29. Equations 22-24 basically tell us that Q and P – which obey the Heisenberg-Liouville equations automatically – also fit the dynamics at any time "0", and thus that $Tr(\rho(t=0)Q_j(t))$ yields $<\varphi_j(t)>$, using the exact same similarity transform argument that Weinberg invokes.

Notice that we only require that $Q_j(t)$, so defined, has the property that $Tr(\rho(0)Q_j(t))=<\varphi_i(t)>$ for any physically realizable matrix $\rho(0)$. There is no reason why $Q_j(t)$ should have to be unique in having that property. In fact, it is possible to derive alternate dynamics for an alternate $Q_j(t)$ matrix which possesses this same property. This is related to the concept of "null equivalence," which is briefly introduced in a previous paper[2], but not central to the goals of this letter.

IV. IMPLICATIONS AND POSSIBLE EXTENSIONS

These results indicate that there is an exact equivalence between the dynamics of a bosonic QFT, in the 0+1-D case, and the statistical dynamics of the corresponding "classical theory," with some minor adjustment of scalar parameters. The extension to 3+1-D should not be difficult; see Appendix B.

Does this imply that bosonic QFTs are totally equivalent to classical field theory? Bell's Theorem experiments tell us that a complete equivalence is impossible[2]. However, QFT is made up of two major components, two bodies of assumptions which are combined together to generate predictions – quantum dynamics and quantum measurement. Classical field theory was not so explicit in breaking out assumptions about dynamics versus assumptions about measurement, but it, too, made strong apriori assumptions about the role of causality and statistics in measurement which were not derived at all from the assumed Lagrange-Euler equations. Therefore, it makes sense to conclude that the differences between QFT and CFT are mainly due to differences in measurement formalisms and assumptions about measurement, rather than differences in dynamics. Many authors have argued that quantum measurement could in fact be derived somehow from quantum dynamics; if true, this might imply that Lagrangian PDE models, governed by Heisenberg dynamics, would actually obey quantum measurement rules as well.

The classical-QFT equivalence in dynamics probably extends well beyond bosonic field theory as such. LeGuillou et al (hep-th/9602017) claim to provide a general procedure for mapping fermionic QFTs into equivalent bosonic QFTs, based on an extension of a general procedure published by Witten in 1984, which they cite. This suggests that the dynamics of the standard model itself may be equivalent to a bosonic QFT, which in turn is equivalent to a classical model.

This then suggests a possible strategy for developing a finite (not only renormalizable) variation of the standard model, by exploiting the classical-QFT equivalence in dynamics.

Many of the renormalizations required in QFT are based on effects like the infinite self-repulsion of a point charge. The same problems exist in the corresponding classical theories. But in classical field theories such effects can be eliminated by representing particles as extended bodies – more precisely, as true topological solitons[5]. In the electroweak part of the standard model, particles mainly derive their mass from the Higgs field, a bosonic field whose detailed properties are still almost unknown[6]; it would be reasonable, then, to modify the Higgs model by adding a Skyrme term, which would "explain why massive particles exist" (something closely related to explaining the mass!), and would ensure the good behavior of the corresponding classical PDE in tasks such as the prediction of scattering amplitudes. The PDE-QFT equivalence could then be exploited to prove the good behavior of the QFT. (A few auxiliary terms, analogous to Fadeev-Popov or Umezawa terms, may be needed to strictly enforce the topological constraints assumed in such models.)

This is only a possible direction for future research; it is certainly not a proof! However, please recall that when Weinberg and Salam first proposed the electroweak model in the 1960s, they did not have a proof of renormalizability – only a plausible basis for hope. The standard model is renormalizable, despite the pessimistic earlier predictions based on power-counting rules of thumb, because special properties of the model (symmetry) allow good behavior even when power-counting rules are not met. It is reasonable to hope for a similar sort of special situation here as well.

The discussion so far in this section only addresses EWT. But there is no reason to believe it could not be extended to strong nuclear interactions, the domain where Skyrme models have proven most useful in practical applications. (Witten and others presented many intriguing suggestions for how to develop more empirically-grounded models of strong interactions, using topological soliton models, in Chodos et al.[7]) It may even be possible to modify such a finite standard model, by applying the same methods John Wheeler used to convert Maxwell's Laws into an "already unified field theory." Thus there may even be some hope, in the long-term, of achieving finiteness and unification in a field theory which does not ask us to assume the existence of additional, speculative dimensions of space-time.

*author's emails: pwerbos@nsf.gov, werbos@ieee.org

1. S. Weinberg, *The Quantum Theory of Fields* (Cambridge University Press, Cambridge, 1995), ch. 7

2. P.Werbos and L. Dolmatova, quant-ph 0008036

3. S. Rajeev, *Phys. Rev. D*, Vol. 29. No.12, p.2944-2947 (1984).

4. S. Mandelstam, *Phys. Rev D,* Vol. 11, No. 10, p.3026-3030, (1975)..

5. Makhankov, Rybakov and Sanyuk, *The Skyrme Model*, Springer (1994).

6. S. Weinberg, ibid, ch. 21.

7. A. Chodos, E. Hadjimichael and Ch. Tze, eds, *Solitons in Nuclear and Elementary Particle Physics* (World Scientific, Singapore, 1984).

APPENDIX A: RELATION TO METHODS IN QUANT-PH 0008036

The classical density matrix $\rho$ defined above is related to some of the representations given in quant-ph 0008036, but is quite distinct. The relationship is analogous to the relationship between Fourier analysis and Laplace analysis. Fourier analysis, Laplace analysis and classical perturbation theory are not alternative theories of the universe; rather, they are alternative methods for analyzing the behavior of ODE and PDE; they are all sometimes valid and sometimes not, even as applied to well-posed ODE and PDE, depending on the properties of the specific ODE and PDE. There is no apriori reason to expect that every well-posed model of nature must fit any one particular choice of method, such as perturbation analysis; indeed, the great success of nonperturbative methods in recent years demonstrates the value of being open to a variety of methods.

The paper quant-ph 0008036 provides a number of alternative representations or methods which may appear quite complicated at first. This appendix will summarize the two representations used in that paper (**v** and **z**) which are closest in spirit to ρ.

The matrix ρ has been defined here as an object over the usual Fock-Hilbert space of bosonic QFT. That space is essentially just the vector space spanned by the vacuum state |0> and by all states which result when we multiply that state by the usual creation and annihilation operators, $a_1^+,...,a_n^+, a_1,...a_n$ in any sequence. However, instead of representing the statistical moments as a matrix over the conventional Fock space, it is possible to represent them as a vector over an expanded Fock space. We can define new creation and annihilation operators, $b_1^+,...,b_n^+$ and $b_1,...,b_n$, which commute with the usual system of $a_1^+,...,a_n$. The expanded Fock space is the space spanned by the states we get when we multiply any of the "a" or "b" vectors by the vacuum state, in any order.

Let us consider how this works out when we analyze the statistics of a second-order ODE system defined by:

$$L(\underline{\varphi}, \underline{\dot{\varphi}}) = \tfrac{1}{2}\underline{\dot{\varphi}}^2 - \tfrac{1}{2}\sum_{j=1}^{n} w_j^2 \varphi_j^2 - f(\underline{\varphi}) \quad , \tag{25}$$

where f is an analytic function. In CFT, this directly yields the Hamiltonian and Lagrange-Euler equations:

$$H(\underline{\varphi}, \underline{\dot{\varphi}}) = \tfrac{1}{2}\underline{\dot{\varphi}}^2 + \tfrac{1}{2}\sum_{j=1}^{n} w_j^2 \varphi_j^2 + f(\underline{\varphi}) \tag{26}$$

$$\ddot{\varphi}_j = -w_j^2 \varphi_j - \partial_j f(\underline{\varphi}), \qquad j=1,...,n \tag{27}$$

This is actually far more general than it seems, insofar as we can use the components of $\underline{\varphi}$ to represent field values at points in some kind of periodic lattice. Note that we have the freedom to pick the $w_j$ parameters to be anything we like, and modify f accordingly, to arrive at equivalent representations of the same system. Likewise, we can use this to represent either locations or frequency modes in the periodic crystal.

The statistical moments of this system may be represented by use of the following vector over the expanded Fock space, for any pure state:

$$\underline{v}(\underline{\varphi}, \underline{\dot{\varphi}}) = \exp\left( \sum_{j=1}^{n} (w_j \varphi_j a_j^+ + \dot{\varphi}_j b_j^+) \right) |0> \tag{28}$$

A statistical ensemble or probability distribution Pr($\underline{\varphi}$, $\partial_t \underline{\varphi}$) may be represented by:

$$\underline{v} = \int \underline{v}(\underline{\varphi}, \underline{\dot{\varphi}}) \Pr(\underline{\varphi}, \underline{\dot{\varphi}}) d^n \underline{\varphi} \, d^n \underline{\dot{\varphi}} \tag{29}$$

In order to simply or "reify" the analysis, we may define another vector $\underline{z}$ over the expanded Fock space:

$$\underline{z}(\underline{\varphi}, \underline{\dot{\varphi}}) = X \underline{v}(\underline{\varphi}, \underline{\dot{\varphi}}) \quad , \tag{30}$$

where X is the "continuous reification operator" defined by:

$$X = \exp\left(-\frac{\pi}{8}\sum_{j=1}^{n}(a_j a_j + a_j^+ a_j^+ + b_j b_j + b_j^+ b_j^+)\right) \qquad (31)$$

This concept of reification was described at length in quant-ph 0008036, and some of the prior papers it cites. Equation 31 is somewhat more convenient than the specific forms of reification chosen there, for several reasons. This operator has the following really critical properties, which can be verified easily by brute-force algebra, expanding equation 31 as a power series and using well-known properties of commutators:

$X\ a_i\ X^{-1} = (a_i + a_i)/\text{sqrt}(2)$ （32)

$X\ a_i^+\ X^{-1} = (a_i^+ - a_i)/\text{sqrt}(2)$ (33)

The same equations also hold if "a" is replaced everywhere by "b". Again, see quant-ph 0008036 for generalization to the PDE case.

Applying well-known methods of commutator algebra to equation 28, we easily deduce:

$a_j\ \underline{v}\ (\underline{\varphi}, \partial_t \underline{\varphi}) = w_j\ \varphi_j\ \underline{v}(\underline{\varphi}, \partial_t \underline{\varphi})$ (34)

$b_j\ \underline{v}\ (\underline{\varphi}, \partial_t \underline{\varphi}) = (\partial_t \varphi_j)\ \underline{v}(\underline{\varphi}, \partial_t \underline{\varphi})$ (35)

and, in general:

$$f(\{w_i\varphi_i\},\underline{\dot{\varphi}})\underline{v}(\underline{\varphi},\underline{\dot{\varphi}}) = f(\underline{a},\underline{b})\underline{v}(\underline{\varphi},\underline{\dot{\varphi}}) \qquad (36)$$

In this case, the definition of f(**a**, **b**) is unambiguous; it does not require us to specify whether we are using the canonical or normal or other form, because these annihilation operators all commute with each other.

We can derive the dynamics of **v** simply by differentiating equation 28 with respect to time, which yields:

$$\underline{\dot{v}}(\underline{\varphi}, \underline{\dot{\varphi}}) = \partial_t \left( \sum_j (w_j \varphi_j a_j^+ + \dot{\varphi}_j b_j^+) \right) \exp\left( \sum_j (w_j \varphi_j a_j^+ + \dot{\varphi}_j b_j^+) \right) | 0 >$$
$$= \sum_j (w_j \dot{\varphi}_j a_j^+ + \ddot{\varphi}_j b_j^+) \underline{v}(\underline{\varphi}, \underline{\dot{\varphi}})$$
(39)

Exploiting equation 27 and defining

$g_j(\underline{\varphi}) = -\partial_j f(\underline{\varphi})$, (40)

equation 39 reduces to:

$$\underline{\dot{v}}(\underline{\varphi}, \underline{\dot{\varphi}}) = \sum_j \left( w_j \dot{\varphi}_j a_j^+ + b_j^+ (-w_j^2 \varphi_j + g_j(\underline{\varphi})) \right) \underline{v}(\underline{\varphi}, \underline{\dot{\varphi}})$$
(41)

If we substitute in from equations 34 and 35, and then exploit linearity to sum over all states in a statistical ensemble, we may deduce for any physically realizable vector **y** that:

$\partial_t \underline{y} = G_{0,v} \underline{y} + G_{I,v} \underline{y}$ (42)

where:

$$G_{0,v} = \sum_j w_j (a_j^+ b_j - b_j^+ a_j)$$
(43)

and:

$$G_{I,v} = \sum_j b_j^+ g_j \left( \left\langle \frac{a_j}{w_j} \right\rangle \right)$$
(44)

where I apologize for the awkward notation in equation 44, where the angle brackets are used to denote the n-component vector made up of the operators $(a_j/w_j)$.

To derive the dynamics of **z**, for any pure state, we simply insert the definition in equation 30 and exploit the relations in equations 32 to deduce:

$\partial_t \underline{z} = G_{0,z} \underline{z} + G_{I,z} \underline{z}$, (45)

where:

$G_{0,z} = G_{0,v}$ (46)

and

$$G_{I,z} = \frac{1}{\sqrt{2}} \sum_j (b_j^+ - b_j) g_j(\underline{\Phi}) \qquad (47)$$

where I now avoid awkward notation by defining the operator vector $\underline{\Phi}$ as the vector made up of the n components:

$$\Phi_j = \frac{a_j + a_j^+}{w_j \sqrt{2}} \qquad (48)$$

Note that the operator $G_z$ is antiHermitian. Thus the vector $\underline{z}$ obeys a Schrodinger equation with all the standard properties assumed in most theoretical discussions of quantum systems. If quantum measurement could be derived as a consequence of quantum dynamics, in principle, then any classical system in this class should actually be governed by quantum measurement as well; however, in recent discussions with Shih, Dolmatova, Dowling and Strekalov, we have found some evidence suggesting that the real story with quantum measurement is even more complicated than the story given in section 2 of quant-ph 0008036 – more complicated for *any* dynamical system, classical or Heisenberg or Schrodinger.

In the $\underline{v}$ and $\underline{z}$ representations, the energy operators are not the same as the "gain" operators $G_v$ and $G_z$. Let E represent the classical scalar energy of a system as defined by equation 26.

There are two types of energy operator for statistical representations like $\underline{v}$ and $\underline{z}$. I will call these $\lambda$-type energy operators and Q-type energy operators.

H is a $\lambda$-type energy operator for $\underline{v}$ if and only if:

$$H \underline{v} = E \underline{v} \qquad (49)$$

for all pure states $\underline{v}$. H is a Q-type energy operator for $\underline{v}$ if and only if:

$$\underline{v}^H H \underline{v} = E |\underline{v}|^2 \qquad (50)$$

for all pure states $\underline{v}$. The definitions for $\underline{z}$ are the same, with "$\underline{v}$ replaced by "$\underline{z}$".
It is easy to see that all λ-type energy operators are also Q-type operators, but that the converse is not in general true.

From equations 26 and 36, it is easy to deduce that equation 49 is satisfied by the operator:

$$H_v = \tfrac{1}{2} \sum_j \left( b_j b_j + a_j a_j \right) + f(<\frac{a_j}{w_j}>) \qquad (51)$$

where again I apologize for the awkward notation. As before, equation 31 effectively induces a similarity transformation on equation 51, such that we may deduce that the following operator is a λ-type energy operator for $\underline{z}$:

$$H_z = \tfrac{1}{4} \sum_j (b_j + b_j^+)^2 + \tfrac{1}{2} \sum_j \Phi_j^2 + f(\underline{\Phi}) \qquad (52)$$

Note that any statistical ensemble or linear combination of pure states will also be an eigenvector of $H_z$, with the same eigenvalue E, *if* and only if all of the states in the ensemble have the same energy E.

This is clearly getting us very close to the way that QFT calculates spectra!! In fact, $H_z$ is identical to $H_Q$, *except* that $H_Q$ replaces the leading term on the RHS of 52 by "$\pi^2$" term which represents $|\partial_t \underline{\phi}|^2$ in terms of the operators $a_i$ and $a_i^+$ only! If one were to apply this formalism to a simple example, like the sine-Gordon system, one would observe that the kink soliton, in the rest state, has a $G_z$ eigenvalue of zero (is not oscillating!) even though it has an $H_z$ eigenvalue corresponding exactly to its classical energy.

To conclude this appendix, I should add some additional comments on the concept of null-equivalence which was introduced briefly but not exhaustively explored in quant-ph 0008036. That paper introduced some alternative density matrices, such as $\rho_z$, which are quite different from the $\rho$ defined above. The complete analysis requires using a concept of "striated matrix" – essentially, a matrix where elements like $\rho_{i,j}$ match elements like $\rho_{i+j,0}$ as required for any physically realizable matrix. *Any* proposed density matrix is null-equivalent (as defined there) to one and only one striated matrix, which may or may not be physically realizable. A striated matrix is physically realizable as a pure state if and only if it is rank one. When an operator is null-equivalent to a valid $\lambda$-type energy operator, it is itself a valid Q-type energy operator. Again, these comments are intended to fill in the analysis of **v** and **z** representations given in quant-ph 0008036, and should not be taken out of context.

APPENDIX B. CLASSICAL-QFT EQUIVALENCE FOR BOSONIC QFT IN d+1 DIMENSIONS

Quant-ph 0008036 provided excruciating detail for extensions of the **v** and **z** and other representations to the case of d+1 dimensional systems. This appendix will discuss a few key details of how to extend the derivation of Heisenberg dynamics in section III of this paper to the d+1 dimensional case. First let us try to analyze the statistical dynamics of the following class of CFT, expressed in Hamiltonian form:

$$\mathtt{L} = \tfrac{1}{2}\sum_{j=1}^{n}\left(\dot{\varphi}_j \pi_j - \dot{\pi}_j \varphi_j - |\nabla \varphi_j|^2 - m_j^2 \varphi_j^2 \right) - f(\underline{\varphi}, \underline{\pi}, \nabla \underline{\varphi}), \tag{53}$$

where $\underline{\varphi} \in R^n$ and f is an analytic function. In theoretical physics, we usually restrict our attention to the special case where the mathematical vector $\underline{\varphi}$ is actually composed of relativistic scalars, vectors, isovectors, tensors, etc.; however it is easier here to address the general case. Whenever the original Lagrangian is relativistically invariant, then the equivalent QFT will also yield predictions consistent with relativity.

Note that equation 2 has some redundancy in it. Any dynamical system with a Lagrangian of this form can be expressed in many equivalent ways; for example, we can change the "mass" terms $m_j$, and balance those changes by changes in f, to arrive at a "different" Lagrangian which is actually the same functional of the fields. (In principle, we could even make the mass terms functions of $|\mathbf{p}|$, after a Fourier transform; however, I have no need of that here.) There is no one "right" choice between these alternatives; all are valid, so long as they are consistent with later stages of analysis.

CFT then gives us the classical Hamiltonian density and Lagrange-Euler equations:

$$H = \tfrac{1}{2} \sum_{j=1}^{n} \left( |\nabla \varphi_j|^2 + m_j^2 \varphi_j^2 \right) + f(\underline{\varphi}, \underline{\pi}, \nabla \underline{\varphi}) \tag{54}$$

$$\dot{\varphi}_j = \frac{\partial f}{\partial \pi_j} \tag{55}$$

$$\dot{\pi}_j = \Delta \varphi_j - m_j^2 \varphi_j - \frac{\partial f}{\partial \varphi_j} - \left( \nabla \cdot \frac{\delta f}{\delta (\nabla \varphi_j)} \right) \tag{56}$$

Let S(t) denote the state of this dynamical system at any time t; S consists of all of the values of $\underline{\varphi}(\underline{x},t)$ and $\underline{\pi}(\underline{x},t)$ across all points in space $\underline{x}$ at time t. Then the classical density matrix may be defined by the functional integral:

$$\rho = \int \frac{\underline{v}(S)\underline{v}^H(S)}{|\underline{v}(S)|^2} \Pr(S) d^\infty S , \tag{57}$$

where:

$$\underline{v}(S) = \exp\left(c \sum_{j=1}^{n} \int (\theta_j(\underline{p}) + i\tau_j(\underline{p})) a_j^+(\underline{p}) d^d \underline{p} \right) |0\rangle \quad , \tag{58}$$

where d is the number of spatial dimensions (i.e. $\underline{x} \in R^d$) and:

$$\theta_j(\underline{p}) = \sqrt{w_j(\underline{p})} \int e^{-i\underline{p}\cdot\underline{y}} \varphi_j(\underline{y}) d^d \underline{y} \tag{59}$$

$$\tau_j(\underline{p}) = \frac{1}{\sqrt{w_j(\underline{p})}} \int e^{-i\underline{p}\cdot\underline{y}} \pi_j(\underline{y}) d^d \underline{y} \tag{60}$$

$$w_j(\underline{p}) = \sqrt{m_j^2 + |\underline{p}|^2} \tag{61}$$

$$c = \frac{1}{\sqrt{2(2\pi)^d}} \tag{62}$$

It is straightforward but tedious to then prove (for d=3) that:

$$\mathrm{Tr}(\rho \Phi_j(\underline{x})) = <\varphi_j(\underline{x})> \tag{63}$$

$$\mathrm{Tr}(\rho \Pi_j(\underline{x})) = <\pi_j(\underline{x})> \tag{64}$$

where the field operators $\Phi_j$ and $\Pi_j$ are defined *exactly* as in Weinberg, equations 7.2.29 and 7.2.30, which in my notation may be written:

$$\Phi_j(\underline{x}) = \Phi_j^+(\underline{x}) + \Phi_j^-(\underline{x}) \tag{65}$$

$$\Pi_j(\underline{x}) = \Pi_j^+(\underline{x}) + \Pi_j^-(\underline{x}) \tag{66}$$

where:

$$\Phi_j^-(\underline{x}) = \left(\Phi_j^+(\underline{x})\right)^H \tag{67}$$

$$\Pi_j^-(\underline{x}) = \left(\Pi_j^+(\underline{x})\right)^H \tag{68}$$

$$\Phi_j^+(\underline{x}) = c \int \frac{e^{i\underline{p}\cdot\underline{x}} a_j(\underline{p})}{\sqrt{w_j(\underline{p})}} d^d \underline{p} \tag{69}$$

$$\Pi_j^+(\underline{x}) = -ic \int \left(\sqrt{w_j(\underline{p})}\right) e^{i\underline{p}\cdot\underline{x}} a_j(\underline{p}) d^d \underline{p} \tag{70}$$

In actuality, to perform these calculations, I first treated the parameter c in equation c as an unknown parameter, and solved for $\Phi_j^+(\underline{\mathbf{x}})\underline{\mathbf{v}}(S)=(\varphi_j(\underline{\mathbf{x}})+ \text{imaginary})\underline{\mathbf{v}}(S)$, and likewise for $\Pi$. (I found it convenient as well to replace "p" by "q" in equations 58-61, before proceeding. Those definitions could be expressed a bit more elegantly, but the form given here was convenient for derivation.)

With these definitions, one would expect the logic behind equation 19 to go through as before, yielding the exact same equation for the PDE case! The standard commutator relations from equations 10-16 would also go through (adjusted to the PDE case) – but may simply be extracted from standard texts like Weinberg, since these are the standard field operators now. An important example is:

$$\left[\Phi_j(\underline{x}), \int f_n(\underline{\Phi}(\underline{y}),\underline{\Pi}(\underline{y})) d^d \underline{y}\right] = i\delta^d(\underline{x}-\underline{y})\left(\frac{\partial f}{\partial \pi_j}(\underline{\Phi}(\underline{y}),\underline{\Pi}(\underline{y}))\right)_n \tag{71}$$

where once again the "n" subscript refers to the normal product version of the function. The derivation should then go through, without the previous caveats about needing to adjust scalars.